\documentclass[letterpaper,preprintnumbers,english,floats,floatfix,amssymb,prd,superscriptaddress,twocolumn,nofootinbib]{revtex4}
\usepackage{amssymb,amsmath}
\usepackage{epsfig,hyperref}
\usepackage[svgnames]{xcolor}
\usepackage{pgf,tikz}
\usepackage{epstopdf}
\usepackage[british]{babel}
\usepackage{enumerate}
\usepackage{enumitem}
\usepackage{makecell}
\usepackage{booktabs}
\usepackage{graphicx}

\makeatletter
\DeclareRobustCommand*{\bfseries}{%
  \not@math@alphabet\bfseries\mathbf
  \fontseries\bfdefault\selectfont
  \boldmath
}
\makeatother

\def\be{\begin{equation}}
\def\ee{\end{equation}}
\def\beq{\begin{eqnarray}}
\def\eeq{\end{eqnarray}}

\makeatletter

\newcommand{\arXiv}[2][]{\href{http://arxiv.org/abs/#2}{\texttt{arXiv:#2\@ifempty{#1}{}{ [#1]}}}}
\makeatother

\begin{document}
\title{Dynamics near the central singularity in spherical collapse}

\author{Jun-Qi Guo}%
\email{sps{\_}guojq@ujn.edu.cn}
\affiliation{School of Physics and Technology, University of Jinan, Jinan 250022, Shandong, China}


\begin{abstract}
We study the dynamics near the central singularity in spherically symmetric collapse of a massless scalar field toward Schwarzschild black hole formation. The equations of motion take different simplified forms in the early and late stages of the singularity curve. We report some fine structures of the analytic solutions and universal features for the metric functions and matter near the singularity.
\end{abstract}
\maketitle

\section{Introduction\label{sec:introduction}}
Rich, nonlinear dynamics exists in curved, empty spacetime. Achievements have been made in related branches, e.g., gravitational collapse, spacetime singularities and gravitational waves etc~\cite{Wheeler_1962,Scheel:2014hfa,Oppenheimer_1939,Penrose_1965,Price_1972,Poisson_1989,Poisson_1990,Choptuik:1992jv,Christodoulou_1994,Choptuik:2015mma,Brady_1995,
Dafermos_2014,Dafermos:2017dbw,Abbott_2016,Akiyama:2019cqa,Hawking_1973,Joshi_2007,Christodoulou_2008,Belinski_2018}.

Spherical scalar collapse is a very basic model for investigating the nonlinearity of general relativity, and some seminal results have been achieved in this field. In Ref.~\cite{Choptuik:1992jv}, critical phenomena in gravitational collapse were discovered via numerical simulations. Under the condition of continuous self-similarity, a set of naked singularities in spherical scalar collapse were constructed in Ref.~\cite{Christodoulou_1994}. Collapse of a spherical scalar field in a preexisting Reissner-Nordstrom spacetime was simulated in Ref.~\cite{Brady_1995}. It was observed that when the scalar field is sufficiently strong, the null mass-inflation singularity near the Cauchy horizon generally proceeds a central, spacelike singularity inside the black hole core. However, in Ref.~\cite{Dafermos_2014}, it was shown that when the perturbations are small, the singular boundary of the two-ended Reissner-Nordstrom spacetime can be nowhere spacelike.

In this paper, we study the dynamics near the central singularity in spherical scalar collapse toward Schwarzschild black hole formation. Analytic information is crucial for understanding the nature of collapse and spacetime singularities. However, due to the complexity of the Einstein equations, analytic information is usually unavailable and numerical simulations are implemented, instead. The good thing is that gravity near singularities is extremely strong, such that in some circumstances the equations of motion take simplified forms and approximate analytic information on the metric and matter can be extracted.

Some analytic results on the dynamics near the central singularity in Schwarzschild black holes and spherical scalar collapse have been obtained in the literature. Neglecting the backreaction of the scalar field on the geometry and treating the scalar field as a linear perturbation, Doroshkevich and Novikov explored the dynamics near the central singularity in a Schwarzschild black hole, and found that the scalar field diverges logarithmically with the areal radius $r$~\cite{Doroshkevich_1978}. A full asymptotic expansion solution to the wave equation near the singularity in a Schwarzschild black hole was obtained by Fournodavlos and Sbierski~\cite{Fournodavlos:2018lrk}. In this solution, the first two leading terms include the principal logarithmic term and a bounded second order term. Taking into account the backreaction of the scalar field on the geometry and using the simplified assumption of quasi homogeneity of spacetime, Burko obtained a series-expansion solution for the metric and scalar field near the central singularity in spherical scalar collapse in polar coordinates~\cite{Burko:1997xa,Burko:1998az}. The blow-up rates for the Kretschmann scalar were investigated in Refs.~\cite{Christodoulou_1991,An:2020agx,An:2020ydc}. For spherical scalar collapse, via numerical simulation and asymptotic analysis, approximate analytic solutions for the metric, scalar field and Kretschmann scalar near the central singularity were obtained in Refs.~\cite{Guo:2013dha,Guo:2015ssa}. Compared to Refs.~\cite{Burko:1997xa,Burko:1998az}, the variations of some parameters for the metric functions and scalar field along the singularity curve were studied in Ref.~\cite{Guo:2013dha}. In this paper, on the basis of Ref.~\cite{Guo:2013dha}, with improved code and further analysis, we obtain some fine structures of the analytic solutions near the central singularity in spherical scalar collapse. Some differences between the solutions in the early and late stages along the singularity curve are observed. The metric functions and matter near the singularity display some universal features.

The paper is organized as below. The methodology for simulating spherical scalar collapse is depicted in Sec.~\ref{sec:methodology}. We explore the dynamics near the singularity in the late and early stages of collapse in Secs.~\ref{sec:late_stage} and \ref{sec:early_stage}, respectively. The universal features for the solutions are discussed in Sec.~\ref{sec:universal}. The results are summarized in Sec.~\ref{sec:summary}. Throughout the paper, we set $G=c=1$.

\section{Methodology\label{sec:methodology}}
Collapse of a massless scalar field $\phi$ in spherical symmetry is simulated in the coordinates~\cite{Frolov_2004},
\be
ds^{2} = e^{-2\sigma(t,x)}(-dt^2+dx^2)+{r^{2}(t,x)}d\Omega^2.
\label{double_null_metric}
\ee
The energy-momentum tensor for $\phi$ is $T_{\mu\nu}=\phi_{,\mu}\phi_{,\nu}-(1/2)g_{\mu\nu}g^{\alpha\beta}\phi_{,\alpha}\phi_{,\beta}$. Then the equations of motion read
\be r(-r_{,tt}+r_{,xx})-r_{,t}^2+r_{,x}^2 = e^{-2\sigma},\label{equation_r}\ee
\be -\sigma_{,tt}+\sigma_{,xx} + \frac{r_{,tt}-r_{,xx}}{r}+4\pi(\phi_{,t}^2-\phi_{,x}^2)=0,\label{equation_sigma}\ee
\be -\phi_{,tt}+\phi_{,xx}+\frac{2}{r}(-r_{,t}\phi_{,t}+r_{,x}\phi_{,x})=0,\label{equation_phi}\ee
where $(_{,t})$ and $(_{,x})$ denote $\partial/\partial t$ and $\partial/\partial x$, respectively. For numerical stability concern near the center, Eq.~(\ref{equation_r}) is rewritten as~\cite{Frolov_2004}
\be -\eta_{,tt}+\eta_{,xx}=2e^{-2\sigma},\label{equation_eta}\ee
and the term $(r_{,tt}-r_{,xx})/r$ in Eq.~(\ref{equation_sigma}) is replaced by $-2me^{-2\sigma}/r^3$~\cite{Csizmadia:2009dm}, where $\eta\equiv r^2$ and $m$ is the Misner-Sharp mass $m$, $g^{\mu\nu}r_{,\mu}r_{,\nu}{\equiv}1-2m/r$. The dynamics of $m$~\cite{Guo:2018yyt} is described by
\be
m_{,t}=4\pi r^2\cdot e^{2\sigma}\left[-\frac{1}{2}r_{,t}(\phi_{,t}^2+\phi_{,x}^2)+r_{,x}\phi_{,t}\phi_{,x}\right].
\label{dmdt}
\ee

\begin{figure}[t!]
  \begin{tabular}{ccc}
  \includegraphics[width=0.46\textwidth]{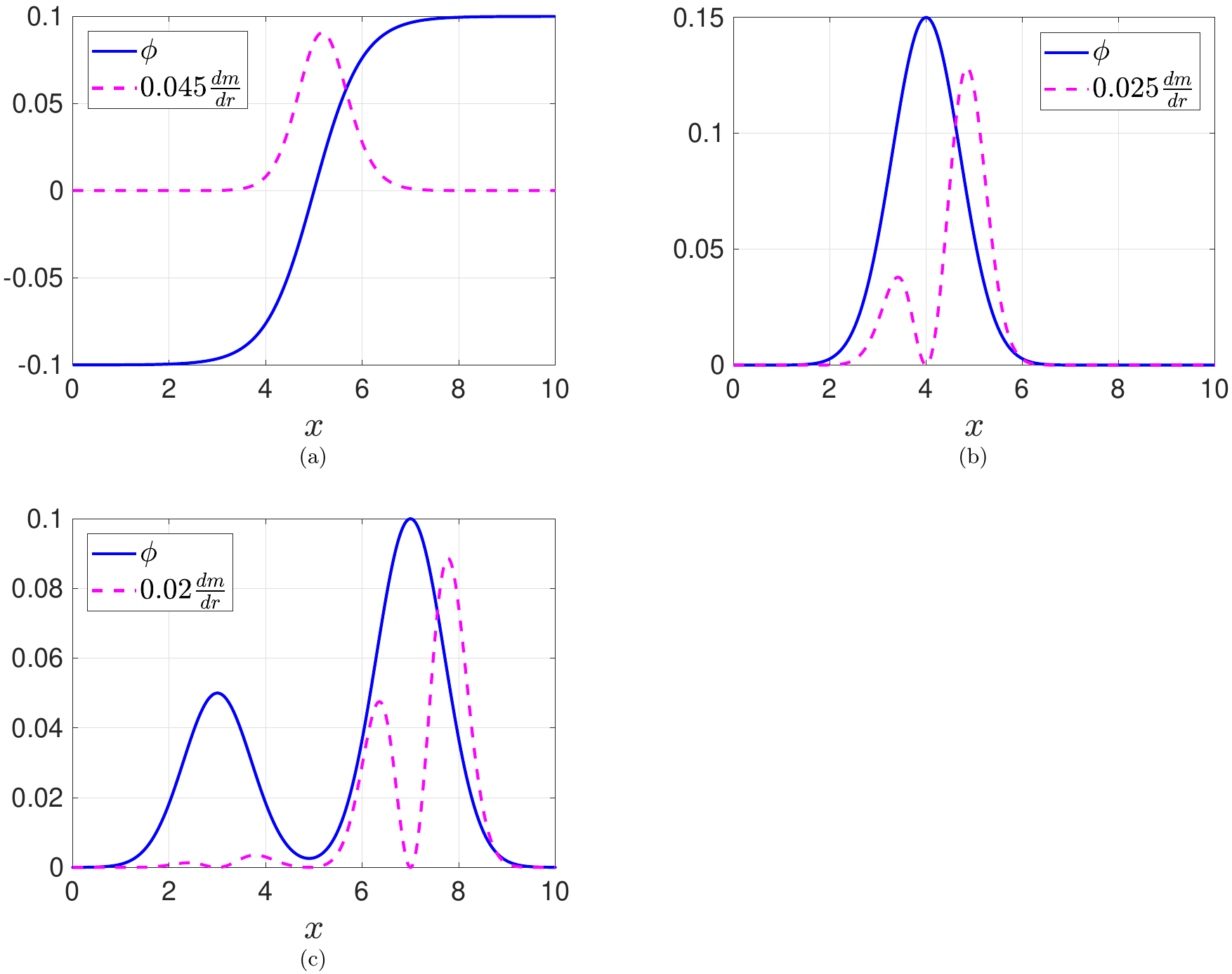}
  \end{tabular}
  \caption{(color online). Initial profiles for the scalar field $\phi$ and the radial mass-energy density $dm/dr(=m_{,x}/r_{,x})$ [see Eq.~(\ref{dmdx})].
  (a) $\phi(t=0,x)=0.1\tanh(x-5)$. (b) $\phi(t=0,x)=0.15e^{-(x-4)^2}$. (c) $\phi(t=0,x)=0.05e^{-(x-3)^2}+0.1e^{-(x-7)^2}$.}
  \label{fig:phi}
\end{figure}

In the simulation, we use the finite-difference method. The numerical setup is basically the same as that in Ref.~\cite{Guo:2018yyt}, and the code is second-order convergent. The initial value for the scalar field is $\phi(t=0,x)=0.1\tanh(x-5)$ [see Fig.~\ref{fig:phi}(a)]. Mesh refinement algorithm~\cite{Garfinkle:1994jb,Guo:2013dha} is implemented in studying the dynamics near the singularity.

\begin{figure*}[t!]
  \begin{tabular}{ccc}
  \includegraphics[width=0.96\textwidth]{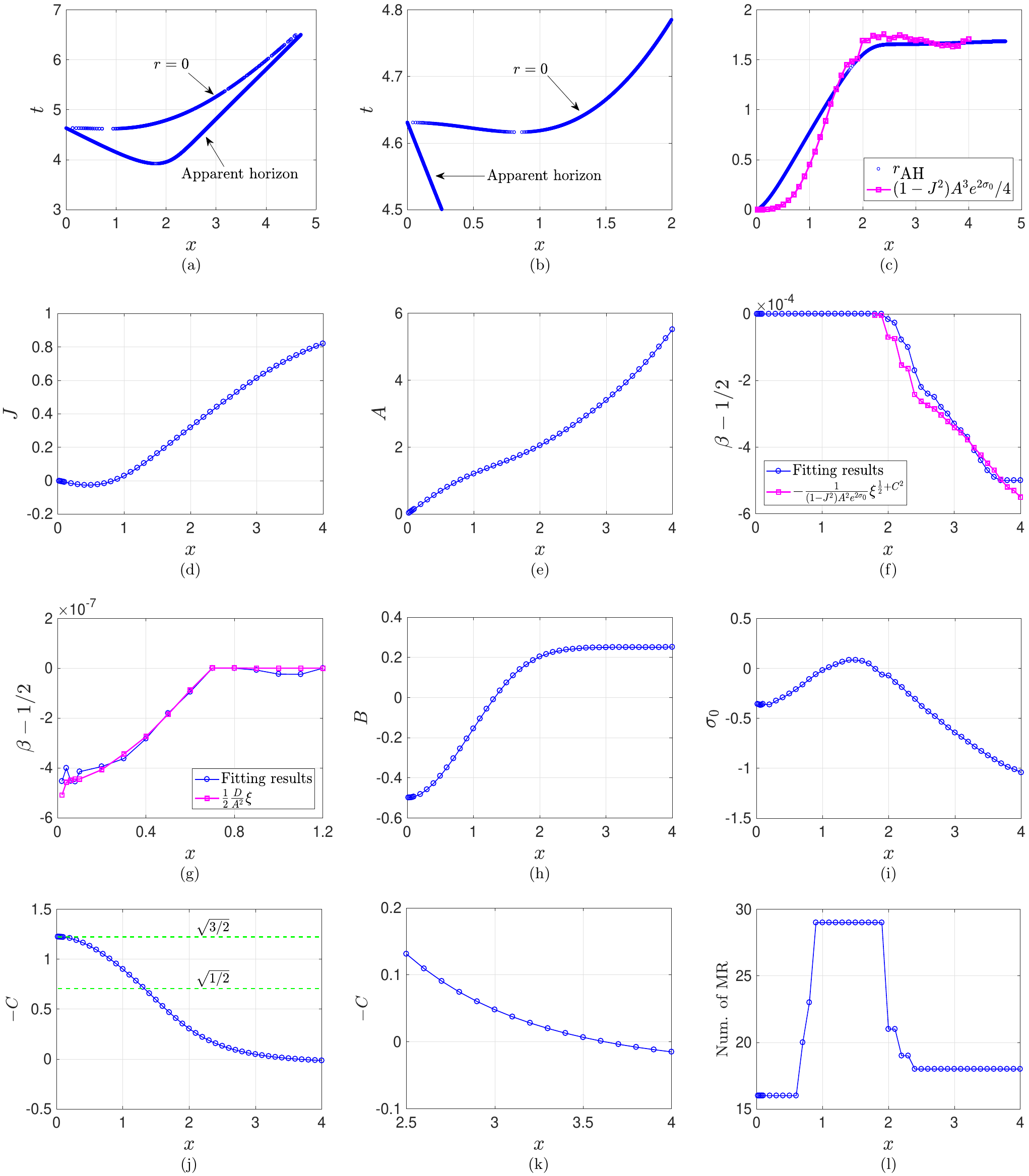}
  \end{tabular}
  \caption{(color online). Numerical results for spherical scalar collapse toward Schwarzschild black hole formation.
  (a) and (b): Apparent horizon and singularity curve.
  (c) Radius of the apparent horizon.
  (d) Slope of the singularity curve, $J$. $|J|$ is obtained from the numerical results of $|r_{,x}/r_{,t}|$. The sign of $J$ is obtained by checking the shape of the singularity curve.
  (e)-(g): Fitting results for $A$ and $\beta$ in Eq.~(\ref{r_asymptotic}), $r\approx A\xi^{\beta}$.
  (h) and (i): Fitting results for $B$ and $\sigma_0$ in Eq.~(\ref{sigma_asymptotic}), $\sigma\approx B\ln\xi+\sigma_0$.
  (j) and (k): Fitting results for $C$ in Eq.~(\ref{phi_asymptotic}), $\sqrt{8\pi}\phi\approx C\ln\xi+\phi_0$.
  (l) Numbers of mesh refinements near the singularity curve, which are also approximately the minimal numbers of mesh refinements needed to make Eqs.~(\ref{ratios}) or (\ref{equation_eta_3}) valid in the simulation. For each round of mesh refinement, we insert one new data point between every two neighbouring grid points in both spatial and temporal directions via interpolation.}
  \label{fig:AH}
\end{figure*}

\section{Result I: Dynamics in the late stage with $|C|<\sqrt{1/2}$\label{sec:late_stage}}
In collapse toward black hole formation, the dynamics at small and large $x$ along the singularity curve of $r=0$ is different. So we separate the singularity curve into early (small $x$) and late (large $x$) stages. In the early (late) stage, the quantity $|C|$ in Eq.~(\ref{phi_asymptotic}) is greater (less) than $\sqrt{1/2}$, and the metric quantity $\sigma$ asymptotes to $+\infty$ ($-\infty$) [see Figs.~\ref{fig:AH}(a), \ref{fig:AH}(h) and \ref{fig:AH}(j) and Eqs.~(\ref{sigma_asymptotic}) and (\ref{B_asymptotic})]. We analyze the dynamics in the late and early stages in this and next sections, respectively.

\begin{figure*}[t!]
  \begin{tabular}{ccc}
  \includegraphics[width=\textwidth]{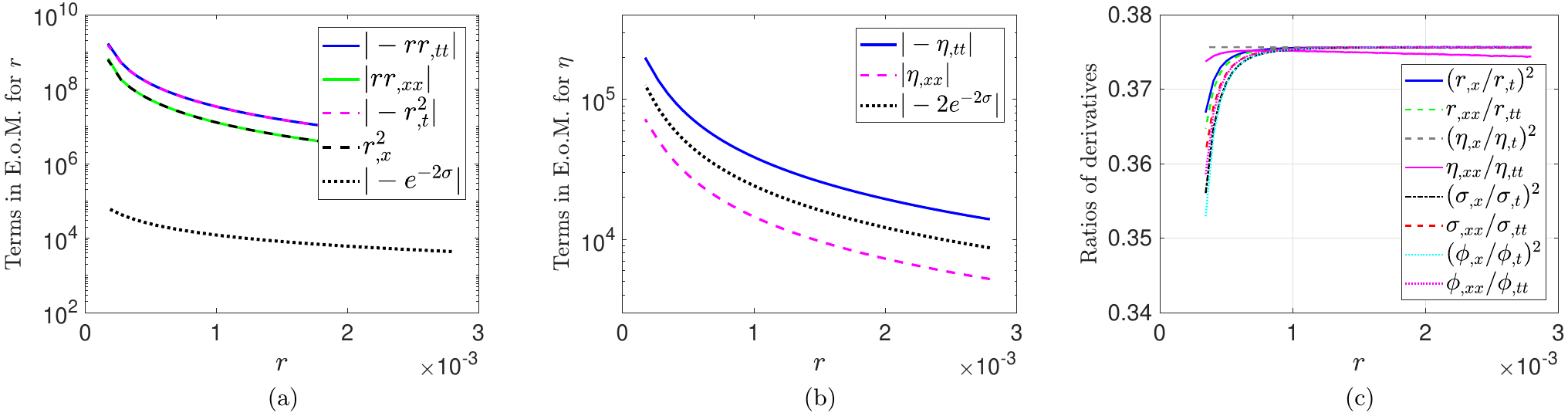}
  \end{tabular}
  \caption{(color online). Numerical results on the late-stage dynamics near the central singularity on the slice of $x=3$.
  (a) Terms in Eq.~(\ref{equation_r}), $r(-r_{,tt}+r_{,xx})-r_{,t}^2+r_{,x}^2 = e^{-2\sigma}$. Near the singularity, $-rr_{,tt}\approx r_{,t}^2$, which yields $r\approx A\xi^{1/2}$.
  (b) Terms in Eq.~(\ref{equation_eta}), $-\eta_{,tt}+\eta_{,xx}=2e^{-2\sigma}$. Near the singularity, the equation is reduced to (\ref{equation_eta_2}): $(-1+J^2)\eta_{,tt}=2e^{-2\sigma}$.
  (c) Ratios between the spatial and temporal derivatives for $r$, $\eta$, $\sigma$ and $\phi$. The ratios are all approximately equal to $J^2$ with $J\approx0.614$.}
  \label{fig:large_x}
\end{figure*}

The numerical results for the apparent horizon and singularity curve $r=0$ of the black hole forming in collapse are plotted in Fig.~\ref{fig:AH}(a). It can be shown that near the central singularity of a Schwarzschild black hole, the ratios between the spatial and temporal derivatives for $r$ and $\eta$ are expressed in terms of the slope $J$ of the singularity curve (see the Appendix). It was also found that in spherical scalar collapse toward black hole formation, near the central singularity, this relation is also true for the quantities of $r$, $\sigma$ and $\phi$~\cite{Guo:2013dha}. In this paper, with more accurate code, we find that such a relation also holds for $\eta$ in the late stage of singularity formation in scalar collapse [see Figs.~\ref{fig:large_x}(b) and \ref{fig:large_x}(c)]. Denoting $Z\in\{r,\eta,\sigma,\phi\}$, we have
\be \left(\frac{Z_{,x}}{Z_{,t}}\right)^2\approx\frac{Z_{,xx}}{Z_{,tt}}\approx J^2.
\label{ratios}
\ee
With Eq.~(\ref{ratios}), Eqs.~(\ref{equation_sigma})-(\ref{equation_eta}) are reduced to
\be
-\sigma_{,tt} + \frac{r_{,tt}}{r} \approx -4\pi\phi_{,t}^2,\label{equation_sigma_Kasner}
\ee
\be
\phi_{,tt}\approx -\frac{2r_{,t}\phi_{,t}}{r},
\label{equation_phi_Kasner}
\ee
\be (-1+J^2)\eta_{,tt}\approx2e^{-2\sigma}.\label{equation_eta_2}\ee

The simplified equations of (\ref{equation_sigma_Kasner})-(\ref{equation_eta_2}), numerical results near the central singularity, and the analysis of dynamics near the singularity for a Schwarzschild black hole together show that, along the slices of $x=\mbox{Constant}$, the quantities of $r$, $\sigma$ and $\phi$ can be well approximated by the following expressions~\cite{Guo:2013dha},
\begin{align}
r&\approx A\xi^{\beta},\label{r_asymptotic}\\
\nonumber\\
\sigma&\approx B\ln\xi+\sigma_{0},\label{sigma_asymptotic}\\
\nonumber\\
\sqrt{8\pi}\phi&\approx C\ln\xi+\phi_0,\label{phi_asymptotic}
\end{align}
where $\xi=t_0-t$ and $t_0$ is the $t$ coordinate on the singularity curve. We respectively fit the numerical results of $r$, $\sigma$ and $\phi$ according to Eqs.~(\ref{r_asymptotic})-(\ref{phi_asymptotic}) with the fitting results being shown in Fig.~\ref{fig:AH}. Substitution of Eqs.~(\ref{r_asymptotic})-(\ref{phi_asymptotic}) into (\ref{equation_sigma_Kasner}) yields~\cite{Guo:2013dha}
\be B\approx\beta(1-\beta)-\frac{1}{2}C^2.\label{B_asymptotic}\ee
As shown in Fig.~\ref{fig:large_x}(a), near the singularity, $-rr_{,tt}\approx r_{,t}^2$, which implies that $\beta\approx1/2$. Considering this result and substituting Eqs.~(\ref{r_asymptotic}), (\ref{sigma_asymptotic}) and (\ref{B_asymptotic}) into (\ref{equation_eta_2}), we arrive at the analytic solution for $\beta$ as described by the first line of Eq.~(\ref{beta_asymptotic_1}).
\be
\begin{split}
\beta&\approx\frac{1}{2}\left[1-\frac{2}{A^{2}(1-J^2)e^{2\sigma_0}}\xi^{\frac{1}{2}+C^2}\right]\\
&\approx\frac{1}{2}\left(1-\frac{A}{4m_0}\xi^{\frac{1}{2}+C^2}\right).
\end{split}
\label{beta_asymptotic_1}
\ee
Here, the second line of the above equation is derived by using the relation regarding the mass of the black hole formed in the late stage of the collapse, $m_0$, given by Eq.~(\ref{m0_BH}). We will postpone the presence of (\ref{m0_BH}) until Sec.~\ref{sec:summary}, where the asymptotical quantities associated with the black hole are investigated, separately.

For large $x$ along the singularity curve, $|C|$ and $\beta$ asymptotes to $0$ and $1/2$, respectively. In this case, with Eq.~(\ref{r_asymptotic}), the second line of (\ref{beta_asymptotic_1}) is simplified as
\be
\beta\approx\frac{1}{2}\left(1-\frac{r}{4m_0}\right).
\ee
Note that the factor of $2/[A^{2}(1-J^2)e^{2\sigma_0}]$ in the first line of (\ref{beta_asymptotic_1}) was omitted in Ref.~\cite{Guo:2013dha}.

We obtain $\beta$ by fitting the numerical results of $r$ vs. $\xi$ according to Eq.~(\ref{r_asymptotic}) for a certain range of $\xi$, $[\Delta \xi, \sim180\Delta\xi]$, where $\Delta\xi$ is the grid distance. We also obtain $\beta$ from the first line of Eq.~(\ref{beta_asymptotic_1}) with $\xi=122\Delta\xi$. As shown in Fig.~\ref{fig:AH}(f), the results by the numerical and analytical approaches match well.

\begin{figure}[t!]
  \begin{tabular}{ccc}
  \includegraphics[width=4.9cm]{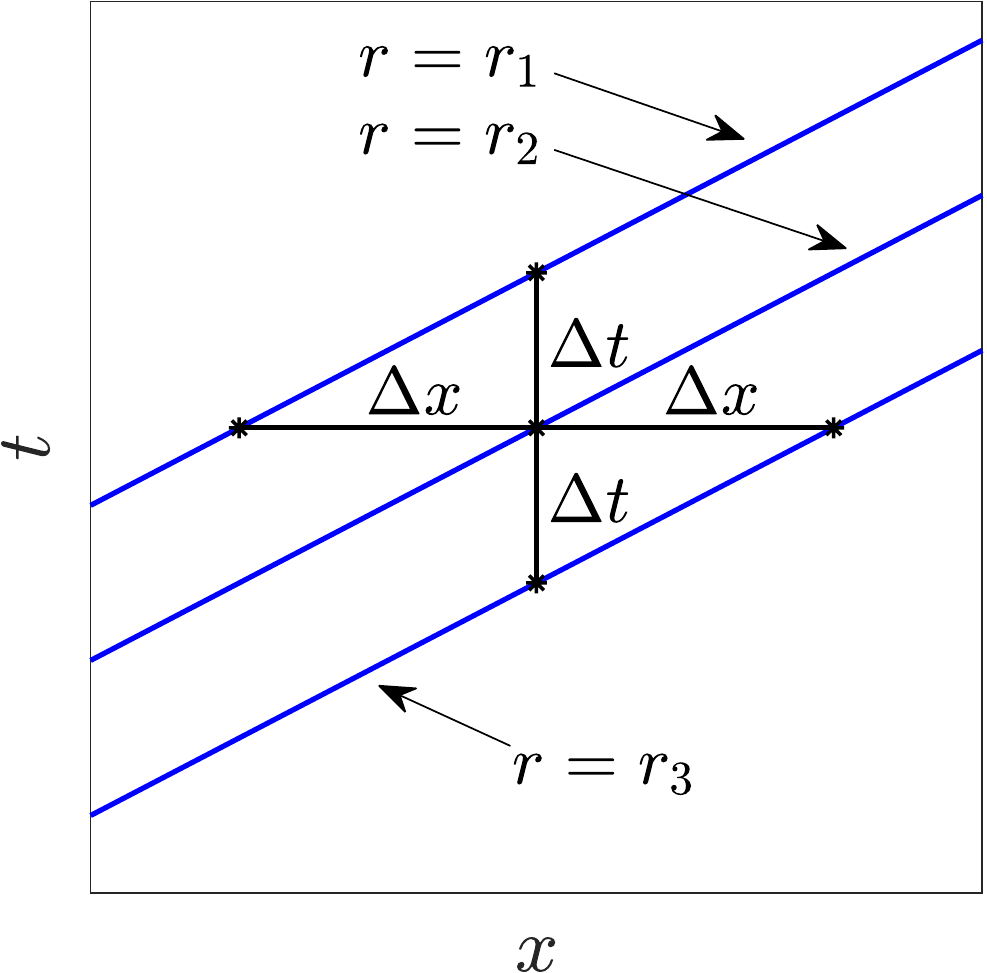}
  \end{tabular}
  \caption{Pictorial illustration of the contour lines of $r$ near the central singularity in a very local region. Denote $J$ as the slope of the contour lines. $J\approx\Delta t/\Delta x\approx r_{,x}/r_{,t}$, and $J^2\approx(\Delta t/\Delta x)^2\approx r_{,xx}/r_{,tt}$.}
  \label{fig:contour}
\end{figure}

\begin{figure*}[t!]
  \begin{tabular}{ccc}
  \includegraphics[width=\textwidth]{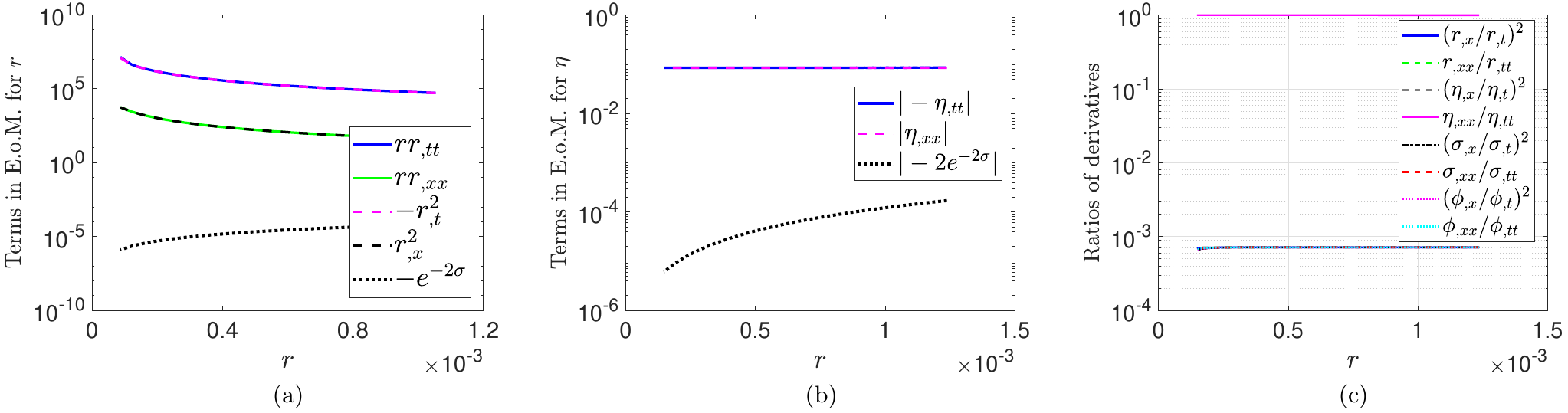}
  \end{tabular}
  \caption{(color online). Numerical results on the early-stage dynamics near the central singularity on the slice of $x=0.5$.
  (a) Terms in Eq.~(\ref{equation_r}), $r(-r_{,tt}+r_{,xx})-r_{,t}^2+r_{,x}^2 = e^{-2\sigma}$. Near the singularity, $-rr_{,tt}\approx r_{,t}^2$, which yields $r\approx A\xi^{1/2}$.
  (b) Terms in Eq.~(\ref{equation_eta}), $-\eta_{,tt}+\eta_{,xx}=2e^{-2\sigma}$. Near the singularity, the equation is reduced to (\ref{equation_eta_3}): $\eta_{,tt}\approx\eta_{,xx}\approx\mbox{Constant}$.
  (c) Ratios between the spatial and temporal derivatives for $r$, $\eta$, $\sigma$ and $\phi$. $\eta_{,xx}/\eta_{,tt}\approx 1$, and all other ratios are approximately equal to $J^2$ with $J\approx-0.0268$.}
  \label{fig:small_x}
\end{figure*}

\begin{figure}[t!]
  \begin{tabular}{ccc}
  \includegraphics[width=5.5cm]{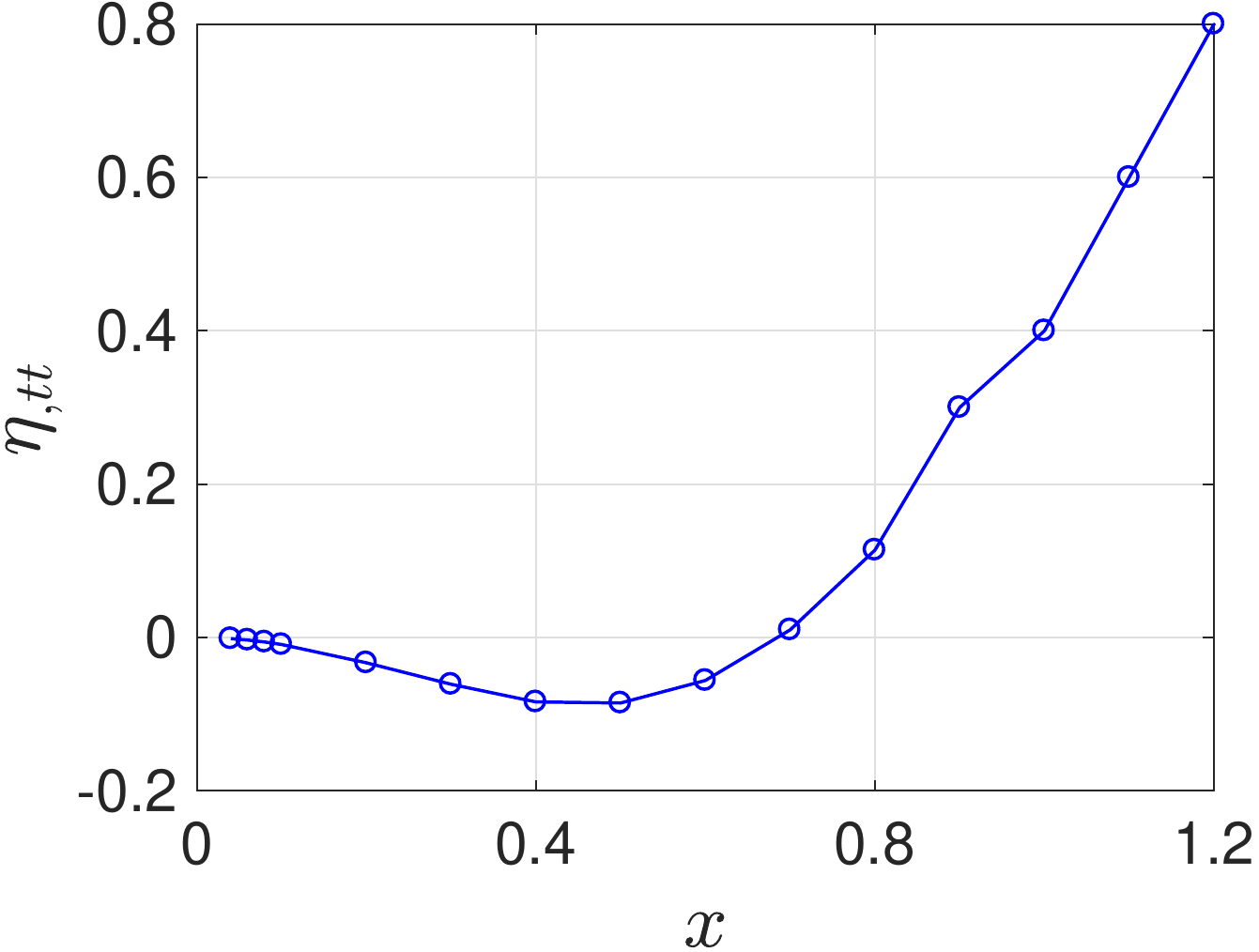}
  \end{tabular}
  \caption{$\eta_{,tt}$ in the early stage of singularity formation.}
  \label{fig:etatt}
\end{figure}

\begin{figure*}[t!]
  \begin{tabular}{ccc}
  \includegraphics[width=0.64\textwidth]{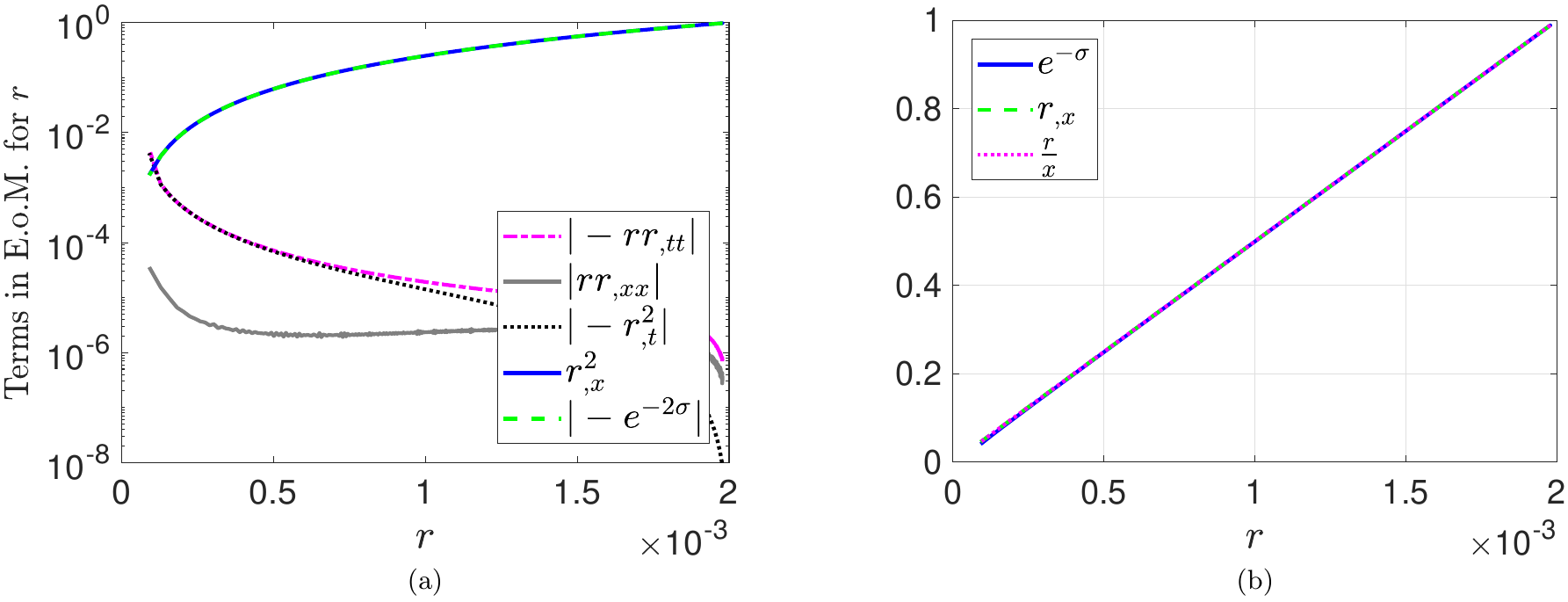}
  \end{tabular}
  \caption{(color online). Conformal flatness of the spacetime near the center in spherical scalar collapse. The results are for the slice of $x=0.002$.
  (a) Terms in Eq.~(\ref{equation_r}), $r(-r_{,tt}+r_{,xx})-r_{,t}^2+r_{,x}^2 = e^{-2\sigma}$. Near the center, $r_{,x}\approx e^{-\sigma}$.
  (b) $e^{-\sigma}\approx r_{,x}\approx r/x$.}
  \label{fig:flat}
\end{figure*}

\begin{figure*}[t!]
  \begin{tabular}{ccc}
  \includegraphics[width=0.61\textwidth]{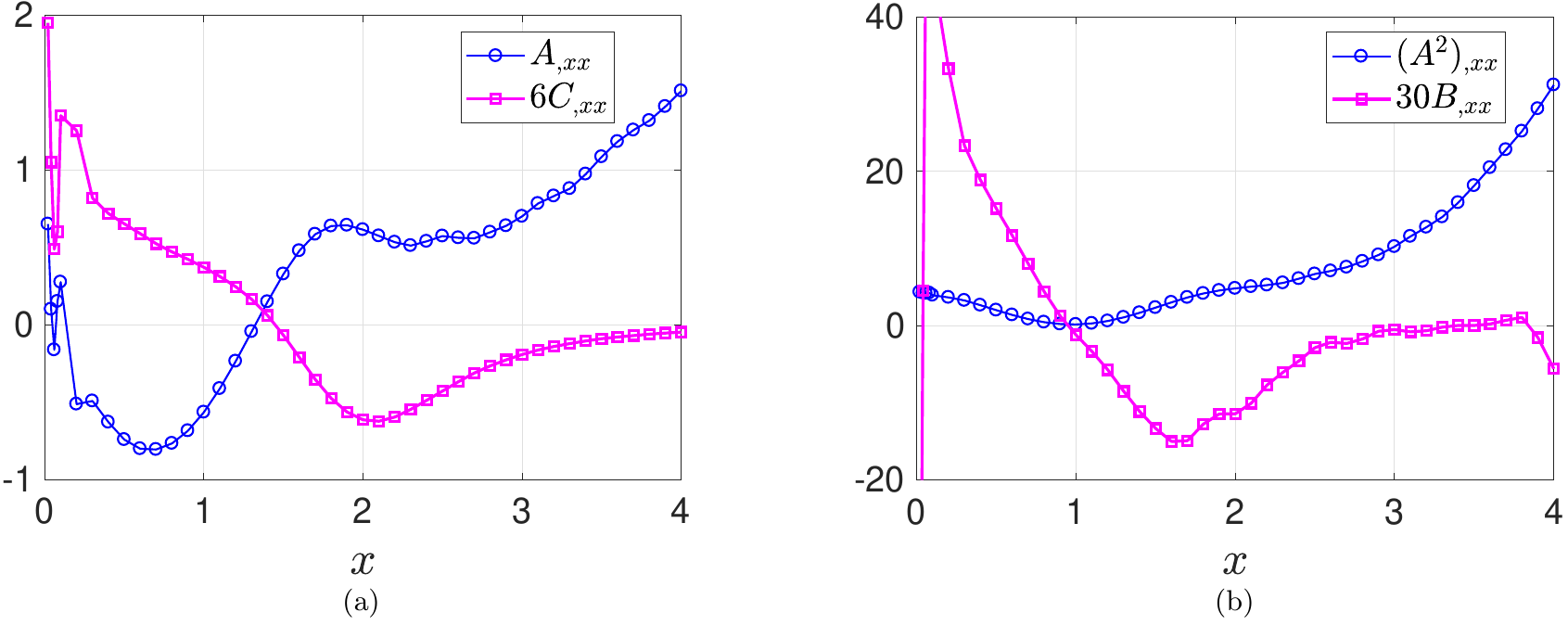}
  \end{tabular}
  \caption{(color online). Some features on the quantities of $A$ in Eq.~(\ref{r_asymptotic}), $B$ in Eq.~(\ref{sigma_asymptotic}) and $C$ in Eq.~(\ref{phi_asymptotic}) along the singularity curve. (a) $A_{,xx}$ vs. $x$ and $6C_{,xx}$ vs. $x$. (b) $(A^2)_{,xx}$ vs. $x$ and $30B_{,xx}$ vs. $x$.}
  \label{fig:Axx}
\end{figure*}

Figure~\ref{fig:AH}(f) seems to show that $\beta$ keeps decreasing as $x$ increases. However, a $\beta$ which is much less than $1/2$ would be very different from the Schwarzschild limit (which is $1/2$). Actually, in Fig.~\ref{fig:AH}(f), $\Delta\xi$ is roughly constant for $2\le x\le4$. For even larger $x$, $\Delta\xi$ needs to be further smaller, such that the region where $\xi\sim\Delta\xi$ on the slice of $x=\mbox{Constant}$ remains inside the horizon and in the vicinity of the singularity. Consequently, near the singularity, $\beta$ will remain close to $1/2$ for large $x$ as expected.

Here we briefly interpret Eq.~(\ref{ratios}). Consider a group of contour lines of $r$ and $\eta$ near the singularity curve in a very local region shown in Fig.~\ref{fig:contour}. From Fig.~\ref{fig:contour}, one can straightforwardly obtain $|r_{,x}/r_{,t}|\approx|\eta_{,x}/\eta_{,t}|\approx J$ and $|r_{,xx}/r_{,tt}|\approx|\eta_{,xx}/\eta_{,tt}|\approx J^2$. Note that, the values of $\sigma$ and $\phi$ described by Eqs.~(\ref{sigma_asymptotic}) and (\ref{phi_asymptotic}) are determined by $\ln\xi$ with $\xi\equiv t_0-t$ and the coefficients $B$ and $C$. Near the singularity, compared to $\ln\xi$, the coefficients $B$ and $C$ change more slowly [see Figs.~\ref{fig:AH}(h) and \ref{fig:AH}(j)]. Then noting the relation (\ref{r_asymptotic}) between $\xi$ and $r$, one obtains that at least in a very local region of the contour lines of $r$, $\sigma$ and $\phi$ also take constant values. So the ratio results of (\ref{ratios}) are also valid for $\sigma$ and $\phi$.

\section{Result II: Dynamics in the early stage with $|C|>\sqrt{1/2}$\label{sec:early_stage}}
\subsection{Dynamics in the early stage with $|C|>\sqrt{1/2}$}
As shown in Fig.~\ref{fig:small_x}(c), in the early stage of singularity formation, Eq.~(\ref{ratios}) remains valid, except for the term of $\eta_{,xx}/\eta_{,tt}$. So Eqs.~(\ref{equation_sigma_Kasner}), (\ref{equation_phi_Kasner}) and (\ref{r_asymptotic})-(\ref{B_asymptotic}) still hold. Therefore, with Eqs.~(\ref{sigma_asymptotic}) and (\ref{B_asymptotic}), one obtains that along the singularity curve, for small $x$ where $|C|>\sqrt{1/2}$, $B$ is negative, and $e^{-2\sigma}$ asymptotes to zero [see Fig.~\ref{fig:small_x}(b)]. Figure~\ref{fig:small_x}(b) also shows that in this case, on slices of $x=\mbox{Constant}$, Eq.~(\ref{equation_eta}) is simplified as
\be \eta_{,tt}\approx\eta_{,xx}\approx\mbox{Constant.}\label{equation_eta_3}\ee
As shown in Fig.~\ref{fig:small_x}(a), in this case, there is also $-rr_{,tt}\approx r_{,t}^2$, which implies that $\beta\approx1/2$. Combining this result and Eqs.~(\ref{r_asymptotic}) and (\ref{equation_eta_3}), we obtain
\be \beta\approx\frac{1}{2}\left(1+\frac{D}{A^2}\xi\right),\label{beta_asymptotic_2}\ee
where $D=\eta_{,tt}\approx{\mbox{Constant}}$ on the slice of $x=\mbox{Constant}$. The fitting results for $\beta$ according to (\ref{r_asymptotic}) and the analytic one by (\ref{beta_asymptotic_2}) match well [see Fig.~\ref{fig:AH}(g)]. The values of $\eta_{,tt}$ are shown in Fig.~\ref{fig:etatt}.

\subsection{Dynamics near $x=r=0$}
The numerical results show that the spacetime near the center before singularity formation is nearly conformally flat~\cite{Guo:2018yyt},
\be e^{-\sigma}\approx r_{,x}\approx\frac{r}{x}.\label{equation_flat}\ee
This is also true for collapse toward black hole formation (see Fig.~\ref{fig:flat}). We interpret Eq.~(\ref{equation_flat}) as below:
\begin{enumerate}[fullwidth,itemindent=0em,label=(\roman*)]
\item We set $r=m\equiv0$ at $x=0$, which is sensible.
With the definition of mass, $g^{\mu\nu}r_{,\mu}r_{,\nu}{\equiv}1-2m/r$, Eq.~(\ref{equation_r}) can be rewritten as
\be -r_{,tt}+r_{,xx} - e^{-2\sigma}\cdot\frac{2m}{r^2}=0.\label{equation_r_2}\ee
The quantity $m_{,x}$ is~\cite{Guo:2018yyt}
\be
m_{,x}=4\pi r^2\cdot e^{2\sigma}\left[\frac{1}{2}r_{,x}(\phi_{,t}^2+\phi_{,x}^2)-r_{,t}\phi_{,t}\phi_{,x}\right].
\label{dmdx}
\ee
Noting that $\phi_{,x}=0$ at $x=r=0$. So near $x=r=0$, $m\propto r^3$, $m/r=0$, and $m/r^{2}=0$.
Considering $r_{,t}=m/r\equiv0$ at $x=r=0$ and $g^{\mu\nu}r_{,\mu}r_{,\nu}{\equiv}1-2m/r$ , we obtain $e^{-\sigma}\approx r_{,x}$ near $x=r=0$.
\item Combining Eq.~(\ref{equation_r_2}) and the result of $r_{,tt}=m/r^2=0$ at $x=r=0$, one arrives at $r_{,xx}\equiv0$, which implies that $r\propto x$ near $x=r=0$. Then we have $r_{,x}\approx r/x$. So Eq.~(\ref{equation_flat}) holds.
\end{enumerate}

When the center just transits into singularity, the spacetime near the center is also conformally flat. In the vicinity of the singularity, the metric is expressed by the Kasner solution~\cite{Guo:2013dha},
\be
ds^2=-d\tau^2+\sum\limits_{i=1}^3 \tau^{2p_i}dx_{i}^{2}.
\label{Kasner_solution}
\ee
$p_1=(-1+2C^2)/(3+2C^2)$, and $p_2=p_3=2/(3+2C^2)$. $q[=4C/(3+2C^2)]$ describes the scalar field's contribution, $\sqrt{8\pi}\phi=q\ln\tau$. The Kasner exponents satisfy $p_1+p_2+p_3=1$ and $p^{2}_{1}+p^{2}_{2}+p^{2}_{3}=1-q^2$. For the conformally flat spacetime, there should be $p_1=p_2=p_3=1/3$ and $|C|\approx\sqrt{3/2}$, which is well supported by Fig.~\ref{fig:AH}(j).

Combining Eq.~(\ref{r_asymptotic}) and $r\propto x$, there are
\be A(x)|_{x\to 0}\approx kx\to 0, \hphantom{dddd}r\approx kx\xi^{\beta}.\label{A_vs_x}\ee
Using Eqs.~(\ref{sigma_asymptotic}), (\ref{B_asymptotic}), (\ref{equation_flat}) and (\ref{A_vs_x}), we have
\be k\approx e^{-\sigma_0},\label{k_vs_sigma}\ee
which is verified by the numerical results: $k\approx1.45$ and $e^{-\sigma_0}\approx1.43$. The first part of Eq.~(\ref{A_vs_x}) is also supported by Fig.~\ref{fig:AH}(e).

Combination of Eqs.~(\ref{ratios}) and (\ref{A_vs_x}) yields
\be |J|_{x\to 0}\approx\bigg|\frac{r_{,x}}{r_{,t}}\bigg|_{x\to 0}\approx \frac{2\xi}{x}.\ee
So near $x=r=0$, the slope $J$ asymptotes to zero, which is confirmed by Figs.~\ref{fig:AH}(b) and \ref{fig:AH}(d).

\subsection{Discussions}
We interpret/discuss the results obtained in this and the previous sections as below:
\begin{enumerate}[fullwidth,itemindent=0em,label=(\roman*)]
\item In spherical scalar collapse, there is a competition between the kinetic energy of the scalar field and its gravitational potential energy. The former tends to disperse the mass-energy of the scalar field to infinity, while the latter tries to trap part of the energy inside a black hole~\cite{Baumgarte_2010}. For regularity concern in Eq.~(\ref{equation_phi}), we set $\phi_{,x}\equiv0$ at $x=r=0$. So under this constraint, $\phi$ is \lq free\rq~to move. Upon singularity formation, the quantity $|C|$ is pushed to its maximum value $\sqrt{3/2}$ by gravity. In this case, the ratios between the three terms in Eq.~(\ref{equation_sigma_Kasner}) are $\sigma_{,tt}:-r_{,tt}/r:4\pi\phi_{,t}^2\approx 2:1:3$. So the term for the kinetic energy of the scalar field, $4\pi\phi_{,t}^2$, dominates over the gravitational term, $-r_{,tt}/r$, and pushes $\sigma$ and $e^{-\sigma}$ to $+\infty$ and $0$, respectively.

  On the other hand, the numerical results show that, for $0<x<1.2$, $|\eta_{,tt}|$ and $|\eta_{,xx}|$ are usually nonzero and bounded, with the exception that $\eta_{,tt}\approx\eta_{,xx}\approx 0$ at $x\approx 0.69$. See Fig.~\ref{fig:etatt}. Then we arrive at Eqs.~(\ref{equation_eta_3}) and (\ref{beta_asymptotic_2}).

  Note that the solution~(\ref{r_asymptotic}) is for the slices of $x=\mbox{Constant}$. Similarly, for the slices of $t=\mbox{Constant}$, the solution for $r$ near the singularity curve can be written as
  \be r\approx H\zeta^{\gamma},\ee
  where $H$ is a function of $t$, $\zeta\equiv|x_0-x|$, $x_0$ the $x$ coordinate on the singularity curve, and $\gamma\approx 1/2$. For $x_0$ close to zero, $\eta_{,xx}$ asymptotes to zero.

\item When the singularity has formed, gravity gradually takes over the dominant role, the scalar field is directly absorbed into the singularity without reflection at the center, and we obtain the results~(\ref{equation_eta_2}) and (\ref{beta_asymptotic_1}). Eventually, the quantity $|C|$ asymptotes to zero.
\end{enumerate}

\begin{figure*}[t!]
  \begin{tabular}{ccc}
  \includegraphics[width=0.98\textwidth]{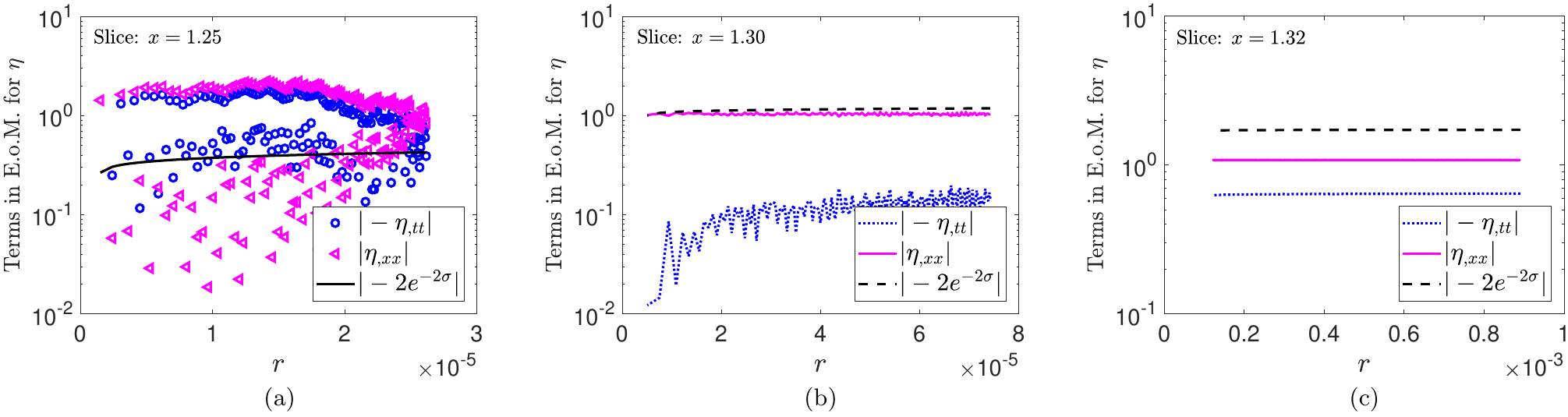}
  \end{tabular}
  \caption{(color online). Behaviors of the terms in Eq.~(\ref{equation_eta}): $-\eta_{,tt}+\eta_{,xx}=2e^{-2\sigma}$. (a) On the slice of $x=1.25$, $\eta_{,tt}\approx\eta_{,xx}$. (b) On the slice of $x=1.30$, $\eta_{,xx}\approx 2e^{-2\sigma}$. (c) On the slice of $x=1.32$, the term of $\eta_{,tt}$ contributes as a minor one.}
  \label{fig:critical}
\end{figure*}

Here we point out/summarize some features at certain points on the singularity curve:
\begin{enumerate}[fullwidth,itemindent=0em,label=(\roman*)]
  \item $x=0$: $A=A_{,xx}=J=0$ [see Figs.~\ref{fig:AH}(b), \ref{fig:AH}(d), \ref{fig:AH}(e) and \ref{fig:Axx}(a)].
  \item $x\approx0.84$: $(A^2)_{,xx}\approx B_{,xx}\approx J\approx0$ [see Figs.~\ref{fig:AH}(b), \ref{fig:AH}(d), \ref{fig:AH}(h) and \ref{fig:Axx}(b)].
  \item $x\approx1.30$: $|C|=\sqrt{1/2}$, and $A_{,xx}\approx C_{,xx}\approx0$ [see Figs.~\ref{fig:AH}(e), \ref{fig:AH}(j) and \ref{fig:Axx}(a)]. At places where $x$ is slightly greater than $1.30$, $J^2\ll1$, then the terms of $\eta_{,tt}$ and $2e^{-2\sigma}$ in Eq.~(\ref{equation_eta}) are dominant, $-\eta_{,tt}\approx2e^{-2\sigma}$. In the early stage where $x<1.30$, the terms of $\eta_{,tt}$ and $\eta_{,xx}$ in Eq.~(\ref{equation_eta}) are dominant, $\eta_{,tt}\approx\eta_{,xx}$. At $x\approx1.30$, $\eta_{,xx}$ and $2e^{-2\sigma}$ are dominant, $\eta_{,xx}\approx2e^{-2\sigma}$ [see Fig.~\ref{fig:critical}(b)].
  \item $x\to\infty$: the Schwarzschild limit.
\end{enumerate}

\begin{figure*}[t!]
  \begin{tabular}{ccc}
  \includegraphics[width=0.94\textwidth]{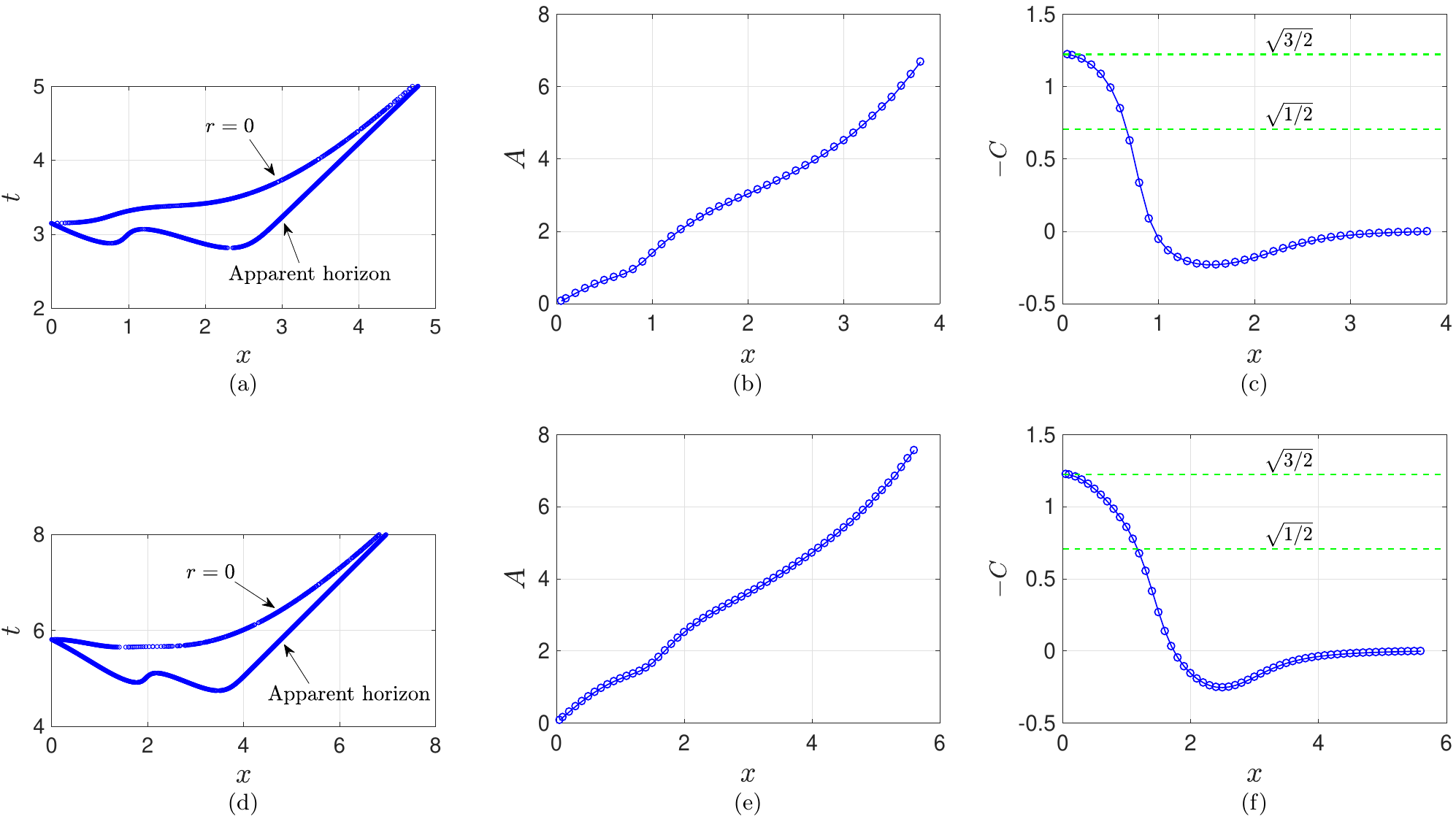}
  \end{tabular}
  \caption{(color online). Numerical results for the second and third sets of initial conditions with $\phi(t=0,x)=0.15e^{-(x-4)^2}$ and $\phi(t=0,x)=0.05e^{-(x-3)^2}+0.1e^{-(x-7)^2}$.
  (a)-(c): Results for the second set of initial conditions.
  (d)-(f): Results for the third set of initial conditions.
  (a) and (d): Apparent horizon and singularity curve.
  (b) and (e): Fitting results for $A$ in Eq.~(\ref{r_asymptotic}), $r\approx A\xi^{\beta}$.
  (c) and (f): Fitting results for $C$ in Eq.~(\ref{phi_asymptotic}), $\sqrt{8\pi}\phi\approx C\ln\xi+\phi_0$.}
  \label{fig:set2}
\end{figure*}

\section{Result III: universal features for the solutions\label{sec:universal}}
In this paper, spherical scalar collapse is simulated with three sets of initial conditions with $\phi(t=0,x)=0.1\tanh(x-5)$, $\phi(t=0,x)=0.15e^{-(x-4)^2}$ and $\phi(t=0,x)=0.05e^{-(x-3)^2}+0.1e^{-(x-7)^2}$ (see Fig.~\ref{fig:phi}). In Ref.~\cite{Guo:2013dha}, the simulation was implemented in $f(R)$ gravity. The fitting results for the parameters related to the metric functions and scalar field along the singularity curve for the four sets of configurations are shown in Figs.~\ref{fig:AH}, \ref{fig:set2}(b) and \ref{fig:set2}(c), \ref{fig:set2}(e) and \ref{fig:set2}(f) in this paper and Fig.~11 in Ref.~\cite{Guo:2013dha}), respectively. It is found that the plot of $A$ vs. $x$ (and of $C$ vs. $x$) has similar shapes. Regarding the plot of $C$ vs. $x$, along the singularity curve from $x=0$ to $x=\infty$, $|C|$ starts from $\sqrt{3/2}$, and asymptotes to zero eventually, and oscillates and decays between the two limits. Therefore, we observe universal features on the dynamics near the central singularity in spherical scalar collapse toward black hole formation.

\section{Summary and discussions\label{sec:summary}}
Considering that in the vicinity of a spacelike singularity, temporal derivatives are much higher than spatial ones, a simplifying assumption of homogeneity was used and a series-expansion solution for the metric functions and scalar field near the central singularity was obtained by Burko~\cite{Burko:1997xa,Burko:1998az}. The solutions in the first-order approximation were rewritten by Hansen \emph{et al.} as below~\cite{Hansen_2005}:
\be ds^{2}=f(r)dr^2+h(r)dt^2+{r^{2}(t,x)}d\Omega^2,\label{metric_polar}\ee
\be f(r)=-(\gamma+2)\frac{1}{2m_0}r^{\beta+2},\label{f_polar}\ee
\be h(r)=2m_{0}Cr^{\gamma},\label{h_polar}\ee
\be \sqrt{8\pi}\phi=\sqrt{2(\gamma+1)}\ln r,\label{phi_polar}\ee
where $m_0$ is the final black hole mass.

Spherical scalar collapse toward black hole formation was also simulated in the double-null coordinates by Burko. It was found that along the singularity, $\gamma$ has a local minimum of $-1$, and the temporal gradient of $\gamma$ is zero at the local minima. So it was expected that at these local minima, at least at the leading order, the singularity is Schwarzschild-like~\cite{Burko:1997xa,Burko:1998az}.

The Schwarzschild-like results are also confirmed in this paper. As shown in Figs.~\ref{fig:AH}(j) and \ref{fig:AH}(k), $-C=\sqrt{3/2}$ at the $x=r=0$ singularity. Along the singularity curve, $-C$ also crosses the line of $C=0$ and is expected to asymptote to zero eventually as $x$ goes to infinity. When $C=0$, the energy-momentum tensor of the scalar field is zero, and the spacetime is vacuum.

The Schwarzschild-limit solution is also verified by the metric component $g_{rr}$. Near the singularity, the Misner-Sharp mass is~\cite{Guo:2015ssa}
\be m\approx\frac{1}{8}(1-J^2)A^{3+2C^2}e^{2\sigma_0}r^{-2C^2}.\label{mass_analytic}\ee
The radius of the apparent horizon, $r=2m_{\scriptsize \mbox{BH}}$, plotted in Fig.~\ref{fig:AH}(c) shows that, for large $x$ where $|C|$ asymptotes to zero, there is
\be m_0\approx\frac{1}{8}(1-J^2)A^{3}e^{2\sigma_0}.\label{m0_BH}\ee
Then using $g^{\mu\nu}r_{,\mu}r_{,\nu}{\equiv}1-2m/r$, the metric component $g_{rr}$ can be written as
\be
g_{rr}=\left(g^{\mu\nu}r_{,\mu}r_{,\nu}\right)^{-1}\approx-\frac{r}{2m_0},
\label{grr}
\ee
which matches well with the Schwarzschild limit. This verifies the usual expectation that in the late stage of collapse, the spacetime near the singularity asymptotes to the Schwarzschild limit. So the no-hair theorem is also valid inside black holes. This is consistent with the conclusions in Refs.~\cite{Doroshkevich_1978,Burko:1997xa,Burko:1998az}.

The dynamics near the central singularity in spherical scalar collapse toward Schwarzschild black hole formation was studied. The equations of motion behave differently in the early and late stages of the singularity curve. Approximate analytic expressions for the metric and matter were obtained. The solutions for the metric functions and matter along the singularity curve show certain universal features.

\section*{Acknowledgments}\small
The author is grateful to Xinliang An, Yun-Kau Lau, Junbin Li, Weiliang Qian, Daoyan Wang, Xiaoning Wu and Lin Zhang for helpful discussions. This work is supported by Shandong Province Natural Science Foundation under grant No.ZR2019MA068.

\appendix*
\section{Derivatives near the singularity curve for Schwarzschild black holes\label{sec:appendix}}
In this Appendix, we derive the spatial and temporal derivatives for $r$ and $\eta(=r^2)$ near the singularity curve for a Schwarzschild black hole in Kruskal coordinates. We will show that the ratios between the spatial and temporal derivatives can be expressed in terms of the slope $J$ of the singularity curve.

For a Schwarzschild black hole in the Kruskal coordinates,
\be
ds^{2} = \frac{32m^3}{r}e^{-r/2m}(-dt^2+dx^2)+{r^{2}}d\Omega^2,
\ee
$r$ is expressed as
\be \frac{r}{2m}=1+W(z),\label{r_W}\ee
where
\be z=\frac{x^2-t^2}{e},\label{z_definition}\ee
and $W$ is the Lambert $W$ function defined by \cite{Corless}
\be Y=W(Y)e^{W(Y)}. \label{lambertW_definition} \ee
$Y$ can be a negative or a complex number. On the hypersurface of $r=\mbox{Constant}$, $z=(x^2-t^2)/e=\mbox{Constant}$. Then, in the 2D space of $(t,x)$, the slope $J$ for the curve of $r=\mbox{Constant}$ can be expressed as
\be J\equiv\frac{dt}{dx}=\frac{x}{t}.\ee

From Eq.~(\ref{lambertW_definition}), one obtains
\begin{align}
\frac{dW}{dz}&=\frac{W}{z(1+W)}, \hphantom{dddddd} \mbox{for } z\neq \left\{0, -\frac{1}{e}\right\}, \label{W_z} \\
\nonumber\\
\frac{d^{2}W}{dz^2}&=-\frac{W^{2}(2+W)}{z^2(1+W)^{3}}, \hphantom{ddi} \mbox{for } z\neq \left\{0, -\frac{1}{e}\right\}.\label{W_zz}
\end{align}
Then with Eqs.~(\ref{r_W}), (\ref{W_z}) and (\ref{W_zz}), one obtains
\be \frac{r_{,x}}{2m}=\frac{dW}{dz}\cdot \frac{2x}{e},\label{r_x}\ee
\be \frac{r_{,xx}}{2m}=\frac{d^{2}W}{dz^2}\left(\frac{2x}{e}\right)^{2}+\frac{dW}{dz}\cdot \frac{2}{e}.\label{r_xx}\ee
Near the singularity curve, $z[=(x^2-t^2)/e]$ approaches $-1/e$, and $W$ asymptotes to $-1$. Then Eq.~(\ref{r_xx}) can be approximated as below,
\be \frac{r_{,xx}}{2m}\approx -\frac{4x^2}{(1+W)^3} \approx \frac{d^{2}W}{dz^2}\left(\frac{2x}{e}\right)^{2}. \label{r_xx_v2}\ee
Similarly, the first- and second-order derivatives of $r$ with respect to $t$ can be written as
\be \frac{r_{,t}}{2m}=-\frac{W}{z(1+W)}\cdot\frac{2t}{e},\label{r_t}\ee
\be \frac{r_{,tt}}{2m}\approx -\frac{4t^2}{(1+W)^3}\approx \frac{d^{2}W}{dz^2}\left(\frac{2t}{e}\right)^{2}. \label{r_tt}\ee

With Eqs.~(\ref{r_x}) and (\ref{r_xx_v2})-(\ref{r_tt}), one obtains
\be \frac{r_{,x}}{r_{,t}}=-\frac{x}{t}=-J,\label{ratio_rx_rt}\ee
\be \frac{r_{,xx}}{r_{,tt}}\approx\left(\frac{x}{t}\right)^2=J^2.\label{ratio_rxx_rtt}\ee
Using $\eta=r^2$ and Eq.~(\ref{ratio_rx_rt}), there is obviously
\be \frac{\eta_{,x}}{\eta_{,t}}=\frac{r_{,x}}{r_{,t}}=-J.\label{ratio_etax_etat}\ee
Combining Eqs.~(\ref{r_W}) and (\ref{W_z})-(\ref{r_xx}), we arrive at
\be \frac{\eta_{,xx}}{8m^2}=\frac{rr_{,xx}+r_{,x}^2}{4m^2}=-\frac{8mW^2}{e^{2}z^2}\frac{x^2}{r}+\frac{2W}{ez}.\label{eta_xx}\ee
Similarly, we obtain
\be \frac{\eta_{,tt}}{8m^2}=\frac{rr_{,tt}+r_{,t}^2}{4m^2}=-\frac{8mW^2}{e^{2}z^2}\frac{t^2}{r}-\frac{2W}{ez}.\label{eta_tt}\ee
Then, near the singularity curve, there are
\be \frac{\eta_{,xx}}{8m^2}\approx-8m\frac{x^2}{r},
\hphantom{dddd}\frac{\eta_{,tt}}{8m^2}\approx-8m\frac{t^2}{r},\ee
\be \frac{\eta_{,xx}}{\eta_{,tt}}\approx\left(\frac{x}{t}\right)^2=J^2.\ee

Combining Eqs.~(\ref{r_W})-(\ref{lambertW_definition}), (\ref{eta_xx}) and (\ref{eta_tt}), we have
\be \frac{-rr_{,tt}-r_{,t}^2+rr_{,xx}+r_{,x}^2}{4m^2}=\frac{8mW}{ezr}=\frac{8m}{r}e^{-r/2m}.\ee
So as expected, with Eq.~(\ref{equation_r}), there is
\be e^{-2\sigma}=\frac{32m^3}{r}e^{-r/2m}.\ee
%


\end{document}